 \documentclass[12pt]{article}
 \usepackage{amssymb}

 \def\barr{\left(\begin{array}}
 \def\earr{\end{array}\right)}

 \def\barr{\left(\begin{array}}
 \def\earr{\end{array}\right)}
 \def\beq#1{\begin{equation}\label{#1}}
 \def\eeq{\end{equation}}
 \def\ber#1{\begin{eqnarray}\label{#1} \nqq}
 \def\eer{\end{eqnarray}}

 \newcommand{\bear}[1]{\begin{eqnarray}\label{#1}}
 \newcommand{\ear}{\end{eqnarray}}

 \catcode`\@=11 \@addtoreset{equation}{section}\catcode`\@=12

 \newcommand{\R}{ {\mathbb R} }

 \newcommand{\fnm}{\footnotemark}
 \newcommand{\fnt}{\footnotetext}
 \newcommand{\np}{ {\newpage } }
 
 \newcommand{\sign}{ \mbox{\rm sign} }
 \newcommand{\e}{ \mbox{\rm e} }

\begin{document}


\begin{center} \large \bf

 Non-singular solutions in multidimensional
    model with scalar fields and exponential potential

\end{center}

\vspace{0.63truecm}

\bigskip

\begin{center}

 \normalsize

   J.-M. Alimi\fnm[1]\fnt[1]{jean-michel.alimi@obspm.fr},

 \bigskip

  Laboratoire de l'Univers et de ses Theories
  CNRS UMR8102,  Observatoire de Paris
  92195, Meudon Cedex, France,

 \bigskip

   V.D. Ivashchuk\fnm[2]\fnt[2]{rusgs@phys.msu.ru}
   and
   V.N. Melnikov\fnm[3]\fnt[3]{melnikov@phys.msu.ru},

 \bigskip

  Center for Gravitation and Fundamental Metrology,
  VNIIMS and  Institute of Gravitation and Cosmology,
  Peoples' Friendship University of Russia,
  3-1 M. Ulyanovoy Str., Moscow, 119313, Russia

 \end{center}

 \bigskip

 \begin{abstract}

 Using developed earlier our  methods for multidimensional models
  \cite{M1,M2,M3} a  family of  cosmological-type solutions
  in $D$-dimensional model with  two
  sets of scalar  fields  $\vec{\varphi}$ and $\vec{\psi}$
  and exponential potential depending  upon $\vec{\varphi}$ is considered.
  The solutions are defined on a product of $n$ Ricci-flat spaces.
  The fields from $\vec{\varphi}$ have  positive kinetic terms and
  $\vec{\psi}$ are "phantom" fields with  negative kinetic terms.
  For  vector coupling constant obeying $0< \vec{\lambda}^2 < (D-1)/(D-2)$
  a subclass of non-singular solutions is singled out. The solutions
  from this subclass are regular for all values of
  synchronous  "time"  $\tau \in ( -  \infty, + \infty )$.
  For  $\vec{\lambda}^2 < 1/(D-2)$ we get an asymptotically accelerated and
  isotropic expansion for large values of $\tau$.

 \end{abstract}

  \np

  \section{Introduction}

 Recently,  the discovery of the cosmic acceleration
 \cite{Riess,Perl} stimulated a lot of papers, on multidimensional
 cosmology in particular, with the aim to explain this phenomenon
 using certain
 multidimensional models, e.g. those of superstring or supergravity
 origin (see \cite{Town,IMS,Chen1,Ohta,IMS1,IMK}, and refs. therein).

 At present rather popular models are those with  multiple
 exponential potential of the scalar fields
 (see \cite{RP,CopLW,BCopN,StarSah} and refs. therein).
 Potentials of such type  arise naturally in certain
 supergravitational models \cite{Town}
 and in sigma-models \cite{IMsigma1,IMC},
 related to configurations with $p$-branes (for a review see also
  \cite{IMSam,IMtop,IMerice}).

  Here we are interested in non-singular solutions
  (e.g. with a bounce) that appear in certain cosmological models, see
  \cite{AWands,BrF} and refs. therein. Such non-singular solutions
  are not new ones.
  For pioneering papers with non-singular solutions based on scalar fields
  see \cite{bm69,SM} - for conformal massless scalar field and cosmological
  constant, \cite{ZKol} - for phantom scalar fields, \cite{MelOrl} - for
  conformal Higgs-type scalar field with spontaneous symmetry breaking
  (SSB) back reaction, \cite{Starpol} - for a scalar field with
  polarization of vacuum (PV) effect, \cite{bmedd} - for a scalar field
  with SSB and PV etc. See all these results also in \cite{SM}.

 In this paper we consider the $D$-dimensional model governed by the
 action

 \bear{1.1}
     S =  \int_{M} d^{D}z \sqrt{|g|} \{
     {R}[g] - g^{MN} \partial_{M} \vec{\varphi} \partial_{N}
              \vec{\varphi}
            \\ \nonumber
             + g^{MN} \partial_{M} \vec{\psi} \partial_{N}
            \vec{\psi}   - 2 V(\varphi) \},
 \ear
     with the scalar potential

 \beq{1.2}
     V(\varphi) =  \Lambda \exp(2 \vec{\lambda} \vec{\varphi} ).
 \eeq

  Here   $\vec{\varphi} = (\varphi^1, \dots, \varphi^k)$
         is a set (vector) of scalar fields,
         $\Lambda $ is a constant,
         $\vec{\lambda} = (\lambda_1, \dots, \lambda_k) \in  \R^k$
         is a vector of dilatonic couplings,
         $\vec{\psi}= (\psi^1, \dots, \psi^m) $  is a
         set (vector) of "phantom" scalar fields,
         $g = g_{M N} dx^{M} \otimes dx^{N}$ is metric,
         $|g| = |\det(g_{M N})|$.

 We note that phantom scalar fields (with negative
 kinetic terms)  appear in various multidimensional
 models, e.g. in a field model for $F$-theory
 \cite{KKLP}, in a chain of $B_D$-models
 in dimensions $D \geq 11$  \cite{IMJ}
 ($B_{11}$ corresponds to  $M$-theory \cite{M-th}
  and $B_{12}$ to $F$-theory \cite{Vafa}).

 The article is devoted mainly to cosmological type
 solutions with the vector coupling constant obeying

 \beq{1.3}
      \vec{\lambda}^2 \neq b \equiv \frac{D-1}{D-2},
 \eeq
 which follow from condition of non-zero scalar product
 for certain $U$-vector \cite{IMS}.

 The paper is organized as follows. In Section 2
 a class of solutions on a product of $n$ Ricci-flat
 spaces  for the model  under consideration is described.
 In Section 3 non-singular   solutions
 (e.g. with simple bounce of volume scale factor)
 are singled out.

 \section{\bf  Solutions on product of Ricci-flat spaces}

  We deal with  the cosmological solution
  to field equations in the model (\ref{1.1})-(\ref{1.3})
  that is  the special case of solutions obtained in \cite{IMS}.
  Our main interest here will be the role of phantom scalar fields
  in combination with the usual matter fields.

  These solutions are defined on the manifold
   \beq{2.1}
     M = (u_{-},u_{+})  \times M_{1} \times \ldots \times M_{n}
   \eeq
   and are described by the metric
   \beq{2.2}
      g =  f^{2b h} [ - \e^{2c^0 t+ 2\bar c^0} w du \otimes du
           +   f^{-2 h} \sum_{i = 1}^n
               \e^{2c^i t+ 2 \bar c^i} g^i ].
    \eeq
     and scalar fields: material
    \beq{2.3}
     \vec{\varphi} = -  h \vec{\lambda} \ln |f| +
                        \vec{c}_{\varphi} t + \vec{{\bar c}},
    \eeq
    and phantom
   \beq{2.3a}
     \vec{\psi} =  \vec{c}_{\psi} t  + \vec{{\bar c}}_{\psi}.
    \eeq

  Here and in what follows we  denote
   \beq{2.4}
     h \equiv (\vec{\lambda}^2 -  b)^{-1}
   \eeq
   and $w = \pm 1$.
   The manifolds $(M_i,g^i)$ are Ricci-flat,
   \beq{02.Ric}
    {\rm Ric}[g^i ] = 0,
   \eeq
   and $d_i = {\rm dim} M_i$,
   $i =1, \dots, n$.

      The  function $f$ reads
   \bear{2.5}
     f = R \sinh(\sqrt{C}(u- u_0)), \;  C>0, \; \epsilon<0;
      \\ \label{2.5a}
      R \sin(\sqrt{|C|}(u- u_0)), \; C<0, \; \epsilon<0;
      \\ \label{2.5b}
      R \cosh(\sqrt{C}(u- u_0)), \; C>0, \; \epsilon>0;
      \\ \label{2.5c}
      |2 \Lambda/h|^{1/2}(u- u_0), \; C=0, \; \epsilon < 0,
    \ear
     where $u_0$ and $C$ are constants and
   \bear{2.6}
        R = |2 \Lambda/(h C)|^{1/2},
             \\ \label{2.6a}
                \epsilon = - \sign(\Lambda w h).
    \ear
   satisfy the following constraint relations

 \bear{2.7}
     \sum_{i = 1}^{n} d_i c^i +  \vec{\lambda} \vec{c}_{\varphi} =0,
     \qquad
     \sum_{i = 1}^{n} d_i \bar c^i +
     \vec{\lambda} \vec{\bar c}_{\varphi} = 0,
 \ear
 and
 \beq{2.8}
     C h + (\vec{c}_{\varphi})^2 -  (\vec{c}_{\psi})^2
          + \sum_{i= 1}^n d_i (c^i)^2
          - \left(\sum_{i=1}^n d_i c^i \right)^2 = 0.
 \eeq

 Relations (\ref{2.7}) means that
 integration constants are orthogonal
 to a certain U-vector
 and relation (\ref{2.8}) is equivalent to the  zero-energy
 constraint, see \cite{GIM,BGrIM,GrIM,IMJ}.

 In (\ref{2.2})
    \beq{2.9}
     c^0 = \sum_{i = 1}^{n} d_i c^i, \qquad
    \bar c^0 = \sum_{i = 1}^{n} d_i \bar c^i.
 \eeq

 The solutions under consideration are general
 $D$-dimensional cosmological type solutions
 defined on product of $n$ Ricci-flat spaces
 $D = \sum_{i = 1}^{n} d_i$ with generalized
 (i.e. multidimensional) Bianchi-I
 metrics in the model (\ref{1.1})-(\ref{1.3}). Here
 $u$ is the harmonic time variable. In what follows we put $u > u_0$.

 \section{\bf Non-singular solutions}

    Here we restrict our consideration to
    small dilatonic couplings when
    \beq{3.1}
        w \Lambda < 0, \qquad   0 < \vec{\lambda}^2 < b
                                     = \frac{D-1}{D-2},
    \eeq
  see (\ref{1.2}). Thus, in (\ref{2.6a}) $\epsilon = -1$.
  As we shall see below the inequalities (\ref{3.1}) guarantee
  the existence of non-singular solutions.

  In what follows we investigate the
  behaviour of solutions from Section 2
  for parameters from (\ref{3.1}) using
  the so-called "synchronous" variable.

  We introduce scale factors
  \beq{3.10}
      a_i = f^{h/(D - 2)} \e^{c^i t + \bar c^i},
  \eeq
  $i = 1,\dots, n$. The relation between
  synchronous and harmonic variables
  reads
  \beq{3.11}
      \tau  = \int_{u}^{u_1} du' v(u'),
  \eeq
   where $u_1 > u_0$ and
  \beq{3.12}
       v = a_1 \dots a_n = f^{bh} \e^{c^0 u + \bar c^0},
   \eeq
   is a volume scale factor.

  From   (\ref{3.11})  for
   $u \to u_0 + 0$  we get $\tau \to +\infty$
   for all values of parameters   and
  \beq{3.12a}
      v \sim \tau^{(D-1) \nu}.
  \eeq

   It follows from (\ref{3.11}) that the solution
   is a non-singular one for $\tau \in ( -  \infty, + \infty )$
   only in such three cases
   \bear{3.13}
      (A_{+}) \quad
              c^0  > b|h| \sqrt{C}, \quad  \  C  \geq 0;
              \\ \label{3.14a}
      (A_{0}) \quad
              c^0  = b|h| \sqrt{C}, \quad  \  C  > 0;
              \\ \label{3.14}
      (B) \qquad \qquad \qquad \qquad   \ C < 0.
    \ear

    In the cases $(A_{+})$ and  $(B)$
    we get a bouncing behaviour of volume scale factor $v(\tau)$
    at some point $\tau_{b}$, i.e.
    the function $v(\tau)$ is monotonically
    decreasing in the interval
    $( -  \infty,  \tau_b )$ and monotonically
    increasing in the interval $( \tau_b, + \infty )$.

    In the case $(A_{0})$  the solution is
    non-singular, the function  $v(\tau)$ is monotonically
    increasing to infinity from a non-zero  value.

    For other values of parameters the solution is
    singular in general position (excepting certain
    special cases),  the function $v(\tau)$
    is monotonically increasing to infinity
    from  zero value. This   takes place when either
          (i) $c^0  < b |h| \sqrt{C}$, $C  > 0$
       or
          (ii) $C = 0$,  $c^0 \leq 0$.
    The solution for
    these values of parameters is defined in the
    semi-infinite interval $( \tau_0, + \infty )$.

    The relations (\ref{3.1}) lead to isotropisation of scale factors
    for large values of synchronous variable $\tau$.
    Namely, for $\vec{\lambda}^2 \neq 0$
    we get the following asymptotical relations
    in the limit $\tau \to +\infty$ \cite{IMS}
     \bear{3.2}
      g_{as}=   w d \tau \otimes d\tau
           + \sum_{i = 1}^{n} A_i \tau^{2 \nu} g^i, \\
                 \label{3.3}
      \vec{\varphi}_{as} = - \frac{\vec{\lambda} }{\vec{\lambda}^2}
                        \ln \tau  + \vec{\varphi}_0,
      \\   \label{3.4}   \vec{\psi}_{as} = \vec{\psi}_0,
     \ear
    where
    \beq{3.5}
      \nu = \frac{1}{(D - 2) \vec{\lambda}^2},
    \eeq
  $A_i > 0$ are constant, and $\vec{\varphi}_0^i$,
  $\vec{\psi}_0$ are  constant
  vectors  obeying
   \beq{3.6}
    2 |\Lambda| \exp( 2 \vec{\lambda} \vec{\varphi}_0) =
    |\vec{\lambda}^2 -  b| / (\vec{\lambda}^2)^2.
   \eeq

 For $\vec{\lambda} = 0$ we get for $\tau \to +\infty$ \cite{IMS}

  \bear{3.7}
    g_{as}=  w d\tau \otimes d \tau +
             \sum_{i = 1}^{n} A_i \exp(2 M \tau) g^i,
      \\ \label{3.8}
             \vec{\varphi}_{as}=  \vec{\varphi}_0,
             \qquad \vec{\psi}_{as}=  \vec{\psi}_0
  \ear
   where  $|2 \Lambda| =  (D-2)(D-1) M^2$
   (in agreement with the case of one scalar field
    from \cite{IM-tmf,BIMZ}.)

 Accelerated expansion in the limit
 $\tau \to +\infty$  takes place for
  \beq{3.9}
      \lambda^2 < \frac{1}{D-2}
  \eeq
  (In (\ref{3.7}) should be $M > 0$ for $\lambda = 0$.)

 \section{\bf The role of scalar charges}

    Here we  show that non-singular solutions
    defined for all $\tau \in \R$ exist only for
    large enough value of (vector) phantom charge squared
    $(\vec{c}_{\psi})^2$.

    Let us introduce anisotropy parameters
      $\tilde c^i$   by the following relations
    \beq{3.19}
     c^i = - \frac{\vec{\lambda}\vec{c}_{\varphi}}{D-1} + \tilde c^i,
    \eeq
    $i = 1,\dots,n$. We get from (\ref{2.9})
    \beq{3.20}
     \sum_{i = 1}^{n} d_i \tilde c^i = 0.
    \eeq

  In terms of these parameters  the main relations
  on parameters (\ref{2.8})
  may be rewritten as following
   \bear{3.21}
     C h +  (\vec{c}_{\varphi}^{~par})^2
      - \frac{1}{b} (\vec{\lambda}\vec{c}_{\varphi})^2
      \\ \nonumber
      = (\vec{c}_{\psi})^2  - (\vec{c}_{\varphi}^{~ort})^2
       -  \sum_{i= 1}^n d_i (\tilde c^i)^2 \equiv Q,
   \ear
  It follows also from (\ref{3.13})  and  (\ref{3.14})
  that
    \beq{3.22}
     c^0 = -  \vec{\lambda} \vec{c}_{\varphi}  \geq
      b h \sqrt{C}, \quad
        {\rm for}  \  C  \geq 0
   \eeq
  and, hence,
    \beq{3.22a}
     (c^0)^2  \geq b^2 h^2 C
    \eeq
  for all non-singular cases $(A_{+})$, $(A_{0})$ and $(B)$.

  Here we use decomposition of the vector of scalar charges
  into a sum of two terms orthogonal and parallel to
  $\vec{\lambda}$, respectively,
     \bear{3.21a}
     \vec{c}_{\varphi} =
     \vec{c}_{\varphi}^{~ort} + \vec{c}_{\varphi}^{~par},
      \\          \label{3.21b}
       \vec{c}_{\varphi}^{~par} =
       (\vec{c}_{\varphi} \vec{\lambda})
       \frac{\vec{\lambda}}{\vec{\lambda}^2},
      \\          \label{3.21c}
       \vec{c}_{\varphi}^{~ort} = \vec{c}_{\varphi}
       - (\vec{c}_{\varphi} \vec{\lambda})
       \frac{\vec{\lambda}}{\vec{\lambda}^2}.
   \ear

   It is clear that
   \bear{3.21d}
     \vec{c}_{\varphi}^{~2} =
     (\vec{c}_{\varphi}^{~ort}) + (\vec{c}_{\varphi}^{~par})^2.
   \ear

   We call $Q$ the "main" parameter of the model.
   We show that the necessary
   condition for the existence of non-singular solutions
   (defined for $\tau \in ( -  \infty, + \infty )$ ) is
   the following one
   \beq{3.Q}
      Q \geq 0,
   \eeq

     that means

   \beq{3.M}
     (\vec{c}_{\psi})^2  \geq (\vec{c}_{\varphi}^{~ort})^2
       + \sum_{i= 1}^n d_i (\tilde c^i)^2.
   \eeq

  Indeed, relations   (\ref{3.21}) and  (\ref{3.22}) imply
    either
    \beq{3.23}
     (A) \quad  Q_{-} \leq Q \leq Q_{+},
         \quad  0 \leq C \leq \vec{\lambda}^2 Q_{-},
    \eeq
    or
    \beq{3.25}
     (B) \qquad  Q > Q_{+},  \qquad  C < 0,
    \eeq
    where
    \beq{3.24}
     Q_{-} = \frac{(b - \vec{\lambda}^2)^2}{b^2 \vec{\lambda}^2}
     (\vec{c}_{\varphi} \vec{\lambda})^2,
     \quad
     Q_{+} = \frac{(b - \vec{\lambda}^2)}{b \vec{\lambda}^2}
     (\vec{c}_{\varphi} \vec{\lambda})^2
      \eeq
     ($\vec{\lambda} \neq 0$). Relation $Q_{-} \geq 0$
     implies  $Q \geq 0$.

   Subclasses (A) and (B) differ by  asymptotical relations
   for    $\tau \to  -  \infty$:
   \beq{3.26}
     (A) \qquad v \sim |\tau|.
   \eeq
   and
  \beq{3.27}
     (B) \qquad v \sim |\tau|^{(D-1) \nu}.
  \eeq
   with $\nu$ defined in (\ref{3.5}).

    The first asymptotics corresponds to Kasner-like behaviour
    of scale factors \cite{IMS} in agreement with the
    billiard approach of \cite{IMsbil,IMpbil}. The second
    one is just the same as in the limit $\tau \to  +  \infty$
    (see \ref{3.12a}).  This  follows from  the relations
    $\sin x = \sin (\pi -x) \sim x$ for small $x = (u - u_0)\sqrt{|C|}$
    and  (\ref{3.11}).

   Non-singular solutions
   defined for $\tau \in ( -  \infty, + \infty )$
   are absent for zero phantom charges $\vec{c}_{\psi} = 0$
   even for isotropic case  with  $\tilde c^i = 0$
   (in this case $\vec{c}_{\varphi} = C = 0$ and
   we get a solution with a power-law
   dependent scale factors, see also \cite{IMS}).

   For $\vec{c}_{\psi} \neq 0$
   the anisotropy parameters $\tilde c^i$ and
   orthogonal component  $\vec{c}_{\varphi}^{~ort}$
   should be small enough (in comparison
   with absolute value of phantom vector charge,
   see (\ref{3.M})) for the existence of
   non-singular solutions defined for all values
   of synchronous variable $\tau \in \R$.

   \section{Conclusions}

  In this paper we have considered
  a family of generalized Bianchi-I
  cosmological type solutions defined
  on product of Ricci-flat spaces
  in $D$-dimensional model  (\ref{1.1})-(\ref{1.3})
  with  $k$ ordinary scalar  fields  $\varphi^i$ and
  $m$ phantom scalar fields $\psi^j$. These
  solutions with phantom scalar fields
  are the special case of the solutions obtained
  earlier in \cite{IMS}.

  For the coupling constant obeying $0 < \vec{\lambda}^2 < (D-1)/(D-2)$
  we have singled out a subclass of
  non-singular solutions (e.g. with
  a simple bounce of the volume scale factor).
  Solutions from this subclass are regular for all values of
  synchronous  time $\tau \in ( -  \infty, + \infty )$
  and have an isotropization behaviour for $\tau \to  + \infty $.

  We have found a necessary condition on scalar charges
  and anisotropy parameters  for the existence of such
  non-singular solutions, see (\ref{3.M}).

  For  $\vec{\lambda}^2 < 1/(D-2)$ this expansion is asymptotically
  accelerated one for large values of $\tau$.

   \begin{center}
   {\bf Acknowledgments}
   \end{center}

   This work was supported in part by the Russian Foundation for
   Basic Researchs  and DFG Project  (436 RUS
   113/807/0-1(R)) for V.D.I. and V.N.M. and Universite Paris 7
   for V.N.M.

  \end{document}